\documentstyle[prl,epsfig,aps]{revtex}

\begin{document}

\twocolumn[\hsize\textwidth\columnwidth\hsize\csname@twocolumnfalse\endcsname
 
\title{Time scale separation and heterogeneous off-equilibrium 
dynamics in spin models over random graphs}
\author{A. Barrat\cite{ab} and R. Zecchina\cite{rz}}
\address{Abdus Salam International Center for Theoretical Physics,
Strada Costiera 11, P.O. Box 563, 34100 Trieste (Italy)}

\maketitle

\begin{abstract}
We study analytically and numerically the statics and the off-equilibrium
dynamics of spin models over finitely connected random graphs. We identify a
threshold value for the connectivity beyond which the loop structure of the
graph becomes thermodynamically relevant. Glauber dynamics simulations show
that this loop structure is responsible for the onset of dynamical features of
a local character (dynamical heterogeneities and spontaneous time scale
separation), consistently with previous (experimental and numerical) studies of
glasses and spin glasses in their approach to the low temperature phase.

\end{abstract}


\vskip.5pc]
 
\narrowtext

Among the characteristic features of the equilibrium and
off--equilibrium behavior of complex physical systems such as glasses or
granular materials, two basic ones are the heterogeneities 
occurring in the spatial 
distribution of particles and the time scale separation in
the relaxation processes of the different degrees of freedom
\cite{exp,Angell,ea,lennardjones}.
The issue of building a clear connection between such local structures 
and the global relaxation process is currently an open basic topic.
In this context, simple spin models have played a crucial role,  both for
the analytical studies which have provided a possible mean--field theory 
of the glass transition \cite{thirum87_parisi_verre_mezpar_bocukumez}
and for numerical simulations.
In particular, the study of mean field spin--glasses with infinite
connectivity has allowed to understand some features of both 
the statics \cite{mpv} and dynamics of the glassy phase \cite{reviewdyn}.
However, intrinsic to the above models is the topology of the connections
which cannot account for any local structure, each site being
connected with all others. 

In this Letter, we will discuss the
relationship between the topology of interactions, in the spirit of 
random networks \cite{Phillips,Thorpe}, and the onset of heterogeneous 
glassy dynamics in models which allow some analytical understanding, namely 
interacting spin models defined over random graphs with finite connectivity.
We will provide analytical and numerical results which
clarify some basic differences between infinitely connected  and finitely 
connected mean field spin models and, more interestingly, identify a simple 
link  between the loop structure of the random lattice, the nature of the
couplings and  the glassy heterogeneous dynamics.  
As we shall discuss, in finitely connected mean field models the non--trivial
underlying topological structure of the links is responsible for the 
appearance of
non--equilibrium phenomena of local nature (both in space and time) and the 
numerical results turn out to be in remarkable agreement with those 
of finite dimensional systems.
Interestingly enough, our results appear also to be related 
with the so called ``small-world'' dynamical patterns discussed in ref.
\cite{Watts}, of relevance in contexts different from physics, e.g. 
biological oscillators, neural networks, spatial games and genetic control 
networks: all these problems are indeed defined over random graphs
where the typical separation between two vertices in the graph is much lower 
than in regular graphs, allowing for a quick spread out of dynamical
correlations over the network (see \cite{Watts} and references therein for a 
more detailed discussion).

Given a graph ${\cal G}=(V,E)$, where $V$ is the set of $N$ vertices and
$E$ is the set of bonds joining $K=2$ (graphs) or $K>2$ 
(hyper-graphs) vertices, the spin Hamiltonians take the form 
\begin{equation}
H=\sum_{i_1<...<i_K \in V} J_{i_1,...,i_K} \, S_{i_1} \cdots S_{i_K} \; \; ,
\label{Ham}
\end{equation}
where the indices $i_1,...,i_K$ run over the set $V$ of $N$ vertices, each
vertex $i$ bearing an Ising spin $S_i$,
and the couplings $J_{i_1,...,i_K}$ associated to the random bonds 
assume values of order one (to be compared with $O(1/\sqrt{N})$ as in usual
infinitely connected models).
We consider graphs with finite connectivity, in 
which the notion of distance is simply the minimal number of bonds on a path 
connecting two sites, also referred to as {\sl chemical } distance.

In the study of random (hyper)graphs the control parameter is the average
density of bonds, $\gamma$ (or the average connectivity $C = K! \gamma$).
For densities small enough, the graph consists
of many small connected clusters of size $O(\ln N)$.
If $\gamma$ increases up to the percolation value $\gamma_c$, there
appears a spanning cluster containing a finite fraction of the $N$ sites 
in the limit of large $N$. However, such a spanning cluster can a priori 
have a tree-like structure, for which the randomness of the couplings 
$J_{i_1,...,i_K}=\pm1$ can be eliminated by a gauge transformation on the 
spins (just like in the one-dimensional random bonds Ising model). 
This leads to the definition of a second threshold value for the density
$\gamma$, defined as that critical density $\gamma_0$ at and beyond which
frustration in the system cannot be removed by such a gauge transformation, 
and therefore gives a macroscopic contribution to the thermodynamics,
raising both internal energy and entropy in the system.
Geometrically this threshold corresponds to the appearance of an extensive
number of loops in the 
spanning cluster and it has been named {\sl percolation of order} (PO) 
transition in ref. \cite{BrayFeng}. 
While, for random $K=2$ graphs, the two transitions are known to coincide
\cite{BrayFeng}, here we shall study the $K=3$ hypergraph structure.
While the notion of loops in hypergraphs is rather counter intuitive, the
idea of frustration retains its simple physical interpretation.
The study of the onset of frustration in the ground state phase diagram of the 
associated random $K=3$ spin glass model,
will allow us to show that the percolation transition $\gamma_c$
and the PO transition $\gamma_0$ are well separated.
Interestingly enough, by resorting to extensive Glauber dynamics simulations, 
we shall also show that such a change in the graph structure is responsible 
for the onset of heterogeneous glassy dynamics.

As discussed in refs. \cite{ksat,GMZ}, the most striking geometrical feature
characterizing the ground state phase diagram of frustrated spin models
over finite connectivity random graphs above the PO transition, is that, in 
spite of a finite entropy per site, there  exists a finite 
fraction of spins which is totally constrained, a ``backbone'' that does not 
change from state to state (strongly reminiscent of rigidity percolation 
\cite{rigidity}).
The remaining fraction of spins is weakly constrained  and
accounts for the overall exponential degeneracy of the ground state.

Here we are interested in the dynamical, off--equilibrium, consequences of 
such a structure consisting, as we shall see, in a spontaneous separation of 
weakly constrained spins (i.e. dynamically fast) and strongly constrained ones,
leading to time scale separation and heterogeneous dynamics at sufficiently 
low temperatures.
Such a behavior, by definition typical of glassy systems \cite{Angell}, 
turns out, for $K \geq 3$, to be independent on the frustration of the 
couplings in that the underlying loop structure together with the K--body 
interaction lead to an annealed self--induced ``geometrical'' frustration.

$K=3$ hypergraphs are constructed as follows:
given $N$ sites, we choose at random $\gamma N$ triplets
$(i,j,k)$ for which the couplings $J_{ijk}$ will be
non zero.
For $\gamma$ greater than the percolation value $\gamma_c=1/3$
($\gamma_c=\frac{2}{K(K-1)}$ for K--spin couplings), 
we obtain, as previously explained, a spanning cluster of connected sites,
containing a finite fraction of the $N$ spins in the large $N$ limit, and
many other smaller (order $\log N$) disconnected clusters \cite{fixed_connect}.

The PO transition is simply identified by comparing
the ground state energy of the random $J_{ijk} = \pm 1$ system with
that of the ferromagnetic system defined over the same hypergraphs.
The value of $\gamma$ beyond which the two energies start to deviate
identifies $\gamma_0$. Such a calculation allows to identify a wide 
{\sl gauge} region, $\gamma_c<\gamma<\gamma_0$, where the relevant 
structure of the spanning cluster is tree--like.

Within the so called Replica Symmetric (RS) functional framework of 
diluted spin glasses \cite{diluted}, we find a first order PO
transition at $\gamma_0^{RS} \simeq 0.9$, characterized by a finite backbone
and finite entropy at the threshold. Such a result was derived by adopting the
full RS iterative scheme discussed in ref. \cite{ksat} in the resolution
of the self--consistency equation for the $T=0$ probability distribution 
of the effective local fields $P(h)=2 R(h/2)$ 
(in the notation of ref. \cite{ksat}), which is given by
$
R(y) = \int _{-\infty} ^\infty \frac{dx}{2 \pi} \cos
( x y) \exp \left\{ -9 \gamma (1-\phi[R;x])\right\}
$
where $\phi[R;x]=\int _{-\infty} ^{\infty} dz_1 dz_2 R(z_1) R(z_2)
\cos [ x \, {\rm min}(1,|z_1|,|z_2|)]$.
Looking for solutions of the form
$R(y) = \sum _{\ell = -\infty}^{\infty} r_\ell \; \delta \left(
y - \frac{\ell}{p} \right)$,
where  $1/p$ is the resolution of the field which eventually goes to zero,
one obtains a set of $p+1$ coupled equations in the independent variables
$\{r_\ell\}$ ($\ell=0,1,...,p$).
Once this set has been solved, 
the ground state energy can be easily derived and compared 
with that of the corresponding
ferromagnetic model, which, as expected, is simply proportional to the average 
connectivity and corresponds to the trivial 
$r_\ell=0$ ($\ell \neq 0$) and $r_0=1$ RS solution.
While the exact identification of the threshold value $\gamma_0$
would require a full Replica Symmetry Breaking solution, indeed an open 
problem, the qualitative features of the phase transition are correctly
identified already at the RS level \cite{rsksat}.
In order to check this fact, we have done exhaustive enumeration of finite 
systems with sizes  $N=18,20,...28$ (averaged over $18000,15000,...3000$
samples respectively).
The extrapolated value for the threshold is $\gamma_0 \simeq 0.8-0.9$
(with a value of the backbone at the transition of $\sim 0.2$),
slightly below $\gamma_0^{RS}$ ($\sim .9$ for $p=15$) as expected. 
However, both the nature of the phase transition and the dependence of the 
ground state energy on $\gamma$ for large connectivity are consistent with 
the RS solution.

For brevity, here we do not report explicitly the details of the above
analysis (a similar calculation is thoroughly described in ref. \cite{ksat}),
but rather focus on the dynamical consequences of the 
topological structure arising from the ground state analysis.
Results from extended Glauber dynamics
simulations performed over various graphs and Hamiltonians,
indeed, confirm the appearance of the expected time scale separation and 
heterogeneous structure of the dynamics in the low temperature 
phase, where the systems are out of equilibrium at all
times, and display aging dynamics. The topology of the 
connections rather than the form of the interactions 
appear to be the source of the robustness of the  phenomenon.
Such a feature is well known in models of glassy material\cite{Thorpe}.

The basic tool used to detect aging dynamics is the spin--spin 
correlation function,
$C(t_w+t,t_w) =\frac{1}{N} \sum_{i=1}^N
\overline{ \langle s_i(t_w+t) s_i(t_w) \rangle} $,
where $t_w$ is the so called waiting time and the average is taken
over the spins, the thermal noise ($\langle . \rangle$)
and the disorder (over-line): while, for equilibrium dynamics,
$C(t_w+t,t_w)$ depends only on $t$, 
aging is defined by the fact that $C(t_w+t,t_w)$ depends also on $t_w$
for all times.
To look for heterogeneities in a system, we
have instead to study more local quantities, like local
correlations $C_i(t_w+t,t_w) = \langle s_i(t_w+t) s_i(t_w) \rangle$
(averaged over the initial conditions),
and the individual rates of flipping of the spins.
For each spin, we can register during one Monte Carlo run the number of times
it flips, and deduce the mean time $\tau_i$ between two spin-flips.
Of course, the total running-time $\tau_{max}$ gives an upper cut-off.
We can then look at the distribution of $\tau_i$, averaged over the samples,
$\overline{P(\tau)}$ \cite{note1}.

We have simulated graphs with
either fixed connectivity ($4$ and $6$) or fluctuating
connectivity (values of $C$ ranging from $3$ to
$18$), and either ferromagnetic or random ($\pm 1$) couplings. To compute
$\overline{P(\tau)}$, we used sizes from 500 to 5000 spins and a number of
samples varying from 50 to 100. No relevant finite size effect
was observed, consistently with the self--averaging character
of $P(\tau)$.
The dynamics has been implemented as a Glauber algorithm with random updating
of the spins, with the runs mostly performed up to a time
of $10^6$ Monte Carlo steps per spin.
For consistency, we have also looked at times up to $10^7$ MC
steps per spin for some samples.

\begin{figure}
\centerline{\hbox{
\epsfig{figure=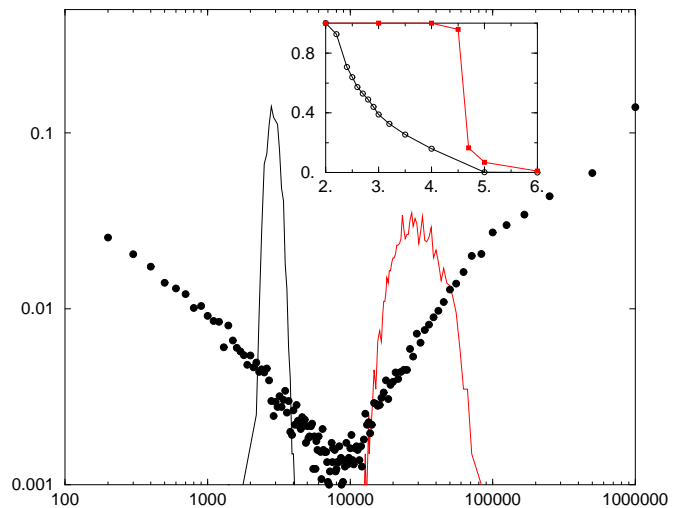,width=7cm,angle=-90}
}}
\caption{Typical examples of the distribution of times $\tau$ between two 
flips of a spin, $\overline{P(\tau)}$, for a hypergraph of
$N=1000$ spins, with fixed connectivity; lines: connectivity $4$,
$\beta = 4$ and $\beta =5$; symbols: connectivity $6$,
$\beta =3$; $t_w=10^4$ MC steps, $\tau_{max}= 10^6$ MC steps. 
Inset: evolution of the fraction of fast spins,
$\int_0^{t^*}  \overline{P(\tau)} d\tau$
with $\beta$, for connectivity $4$ and $6$.}
\label{ptau}
\end{figure}

At high temperature, we find of course a quite simple $\overline{P(\tau)}$, 
peaked at small values of $\tau$, i.e. high flipping rates, for all
connectivities.
However, as the temperature is lowered, we observe very different
behaviors for different mean connectivities. Let us first concentrate
on the case of fixed connectivity. For a random hyper-graph with
connectivity $4$, $\overline{P(\tau)}$ is a smooth function 
peaked around a mean 
value (increasing as $\beta$ is lowered): the dynamics is
homogeneous, and all the spins have more or less the same relaxation
time (with fluctuations). For a fixed connectivity equal to $6$, on the
contrary, we see that small values of $\tau$ still keep a finite weight, while
a second part of the curve, corresponding to large times,
emerges as $T$ decreases. 
This second part appears at temperatures for which aging dynamics
sets in (i.e. at which $C(t_w+t,t_w)$ becomes a function of
both $t$ and $t_w$ for all times), thus signaling the onset of a glassy
regime. The cusp in the curve, around $t^* \approx 10^4$ MC steps per spin, 
shows that a separation of time scales
occurs.
Such a cusp identifies ``fast'' and ``slow'' degrees of freedom,
and persists for a large range of values of $\beta$:
the fraction of fast spins, $\int_0^{t^*} \overline{P(\tau)} d\tau$, 
is a slowly
decreasing function of $\beta$. For the case of connectivity $4$,
$\int_0^{t^*} \overline{P(\tau)} d\tau$ shows instead a sharp 
transition when the
mean value of the relaxation times crosses $t^*$. We show
in figure (\ref{ptau}) the two shapes of $ \overline{P(\tau)}$,
for connectivity $4$ and $6$, and the evolution
of $\int_0^{t^*}  \overline{P(\tau)} d\tau$ with $\beta$. 

If we now consider random hyper-graphs with fluctuating connectivity,
we observe a crossover between the two situations, as $\gamma$ (and
therefore the mean connectivity $C$) is increased. Since the connectivity 
can vary from one site to another, the global $ \overline{P(\tau)}$
can be decomposed in a sum of $ \overline{P_z (\tau)}$,
distributions of times $\tau_i$ restricted to the
sites with connectivity $z$. Then, for 
$\gamma < \gamma_0$, the  $ \overline{P_z (\tau)}$ are smooth functions
peaked around a mean value (evolving with $\beta$), while,
as $\gamma$ grows, the  $ \overline{P_z (\tau)}$
becomes broader and broader, overlap with each other, and exhibits 
cusps \cite{aging}.
The crossover occurs around $\gamma_0$ 
(note that mean connectivity $C=4$ corresponds
to $\gamma < \gamma_0$, and mean connectivity $C=6$ to
 $\gamma >  \gamma_0$), thus indicating that
the loop structure is responsible for the appearance of 
complicated, inhomogeneous dynamics.

\begin{figure}
\centerline{\hbox{
\epsfig{figure=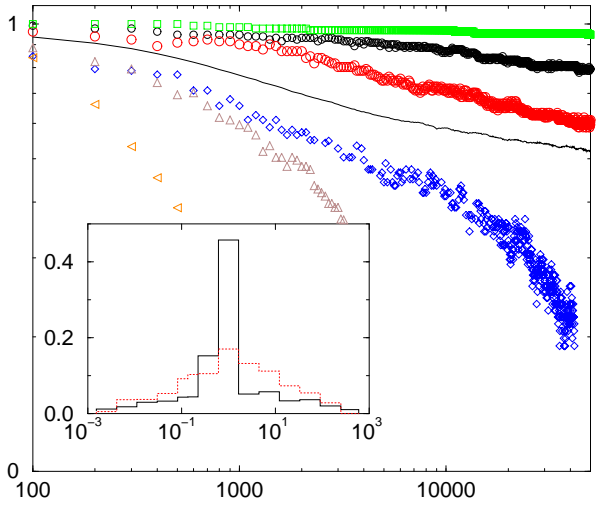,width=8.5cm}
}}
\caption{For one given random hypergraph, local autocorrelation function
$C_i(t_w+t,t_w)$ versus $t$ for various sites $i$ and a given $t_w$.
For some sites $i$ $C_i$ decreases very fast, while for others it
evolves very slowly. The line is the mean
$C(t_w+t,t_w)$.
Here $N=1000$, $\gamma = 1$, $\beta=3.5$,
$t_w=10^4$ MC steps,and the mean is over $300$ runs.
Inset:
histograms of the ratios $\tau_i^{(1)}/\tau_i^{(2)}$ between two runs
with different initial conditions for a random hypergraph with fixed
connectivity (dotted line) and without (full line).}
\label{ci_pt1t2}
\end{figure}

In order to understand the Glauber dynamics mechanisms on a microscopic 
level, we analyze in detail single samples of random hyper-graphs
(with $\gamma > \gamma_0$).
Rather than averaging over disorder, we compare single runs and average
over initial conditions.
Each single run (i.e. initial configuration) leads to
a broad distribution of $\tau_i$. However, two cases may be distinguished.
(I) if the connectivity can fluctuate from site 
to site, $\tau_i$ does not have a strong dependence on the initial conditions:
if we call $\tau_i^{(1)}$ and $\tau_i^{(2)}$ the values of $\tau_i$
for two independent runs, we see in the inset of figure (\ref{ci_pt1t2})
that the histogram of
the ratio $\tau_i^{(1)}/\tau_i^{(2)}$ is sharp and close to one. 
Thus, the broad distributions of
$\tau_i$ lead to a broad distribution of $\langle \tau_i \rangle$, and the
relaxations of the single site correlation functions $C_i(t,t')$ depend
strongly on the site $i$,(see Fig. \ref{ci_pt1t2}). This shows that
the position of the slow and fast degrees of freedom are 
encoded in the topology of the graph.
(II) On the contrary, for random hyper-graphs with uniform connectivity,
the histogram of figure (\ref{ci_pt1t2}) shows that the distribution of
$\tau_i^{(1)}/\tau_i^{(2)}$ is broad (for a given site $i$, $\tau_i$ can vary 
a lot from one run to the other).
It follows that $\langle \tau_i \rangle$ tends to a value independent 
of $i$ in the limit of many runs, and therefore 
$\langle s_i(t) s_i(t') \rangle = \overline{\langle s_i(t) s_i(t') \rangle}$.
The slow or fast character of a spin depends on the initial conditions and 
is induced dynamically. Such a self induced frustration is probably more 
similar to what happens in real glasses.

Surprisingly enough, both for random and constant connectivity hyper-graphs, 
$\langle \tau_i \rangle$ does not depend on the nature of the couplings, 
either ferromagnetic or random.
Once the loop structure of the graphs becomes irregular, frustration is
totally induced by the dynamics through the initial conditions.
For the sake of completeness, we have repeated the simulations on systems
with Gaussian continuous couplings. All the discussed dynamical features
are retained \cite{gaussian}.

As a concluding remark, let us notice that, while the
scope of this study was to show how mean field models could provide  
non--trivial local dynamical information, similar results are found in
finite dimensional Edwards-Anderson spin glasses.
In the latter case the topology of the lattice is fixed and the way of
generating an effective non trivial topology is by randomly adding 
antiferromagnetic bonds to the ordinary Ising model \cite{BrayFeng}.
For the  Edwards-Anderson spin glasses with equally distributed ferro and 
antiferro bonds,
we have also simulated aging dynamics in two and three 
dimensions \cite{ecco}, with the scope of pushing further the study
of \cite{ea}. We indeed observe a similar 
shape for $\overline{P(\tau)}$. 
For ferromagnetic models on regular lattices instead,
the well known coarsening relaxation process occurs: all spins
are equivalent and no rich structure of $ \overline{P(\tau)}$ or
long-lasting spatial heterogeneities can be found. This is also true 
for spin models on quasi-periodic lattices, and on random
graphs (i.e. with $K=2$ instead of $K=3$):
the dynamics shows a similar $\overline{P(\tau)}$, 
with heterogeneities, only for random interactions (Viana-Bray model), 
and not for ferromagnetic ones.

We are most indebted to R. Monasson for valuable comments,
and we thank S. Franz, S. Kirkpatrick, A. Maritan, D. Sherrington and 
A. Vespignani for discussions.

\begin{thebibliography}{99}
\bibitem{fixed_connect}
In order to understand the effects of fluctuations in the site connectivity,
in the simulations we have also considered random graphs with almost
fixed connectivity. Such graphs can be generated
by progressively adding bonds at random and saturating the connectivity
condition for each site, stopping such a process when
the connectivity condition is satisfied for at least $98\%$ of the sites.
\bibitem{note1} Notice that this distribution is self-averaging: its shape is 
the same for every realization of the disorder, which means that the obtained
features are valid for each.
\bibitem{rsksat} In the similar case of the $3-SAT$ model which can be 
studied with very efficient recursive algorithms, the RS threshold
provides an upper bound within $8\%$ of the numerical estimate \cite{ksat}.
\bibitem{aging}
The influence on the waiting time was also investigated. Of course 
$C(t_w+t,t_w)$ depends on the two times, but the
shape of $\overline{P(\tau)}$ is much less affected by $t_w$: the
effect of larger waiting times is just to saturate very slowly the
fraction of frozen spins.
\bibitem{gaussian} 
The only difference is that the distinction between fixed connectivity 
and fixed average connectivity disappears due to the additional fluctuations 
in the values of the couplings.
\bibitem{ecco}
The critical temperature of the 2-dimensional Edwards-Anderson spin glass
is 0, but, for low enough temperature, the system remains out
of equilibrium for very long times, and, for a large
enough system and not too long times, the overall behavior
is the same as the one of a genuine low temperature phase.
\end{thebibliography}
\end{document}